# What is the impact of labor displacement on management consulting services?[1]

A model and its application to the case of consulting firms supporting the German transportation sector

Author: Edouard Ribes, CERNA, Mines ParisTech, France, Email: edouard-augustin.ribes@mines-paristech.fr

## Abstract

Labor displacement off- or nearshore is a performance improvement instrument that currently sparks a lot of interest in the service sector. This article proposes a model to understand the consequences of such a decision on management consulting firms. Its calibration on the market of consulting services for the German transportation industry highlights that, under realistic assumptions, labor displacement translates in price decrease by -0.5% on average per year and that for MC practices to remain competitive/profitable they have to at least increase the amount of work they off/nears shore by +0.7% a year.

---

# 1. Introduction.

The management consulting sector (referred to as MC in the rest of this paper) has expanded tremendously over the past decades (see (Empson, et al. 2015)). This growth has in turn sparked an increase in interest from the academic community across all social sciences. However, according to the recent literature review of (Skjølsvik, Pemer et Løwendahl 2017), several questions remain open with respect to MC firms. One of them pertain to the increase of the competitiveness via the transformation of their service delivery model. A viable instrument to improve performance consists in displacing labor off/near-shore to lower the costs associated to the production of MC services. However, this topic has not been thoroughly discussed in the literature so far. Researchers have indeed merely started to scratch the surface of this type of transformation. If they have recently converged on the fact that these relocations may impact 20% to 50% of the activities associated to the production of consulting services (Jensen et Kletzer 2010), little is known on the consequences of such an evolution for the prices, revenues and profitability of the firms in the MC sector. This gap appears rather surprising given that labor displacement is a well discussed topic when it comes to traditional industrial sectors (such as the manufacturing one).

To address this shortcoming, this paper builds on two strands of the economic and management literature. First, it expands on the very large literature on trade. Displacing labor to improve firm performance is not a new phenomenon but it has mainly been discussed in the context of manufactured goods (Baldwin et Robert-Nicoud 2014). In this set-up, off/near shoring have been shown to lead to price (and thus revenue) reductions on top of obvious cost savings. The results on firm's profitability therefore appear mixed (Arkolakis, et al. 2019) and have triggered intense debates. This article adopts an original point of view by extending the discussion towards considerations of transformation speed. The proposed model indeed shows, that, in order to benefit from trade, a firm must globalize its production engine faster than its competitors. Interestingly, the outcome of this race is an environment consolidated around a few high-performing firms, which echoes the recent macro level findings of (Autor, et al. 2020).

The second contribution of this paper is to extend the growing literature on services firms and their performance improvement efforts. When it comes to labor displacement, the stream has been focused to date on confirming that off/near-shoring is a viable option. This was done by looking at the types of activities that could be done abroad (Ellram, Tate et Billington 2008) and by estimating how much of the production chain of those firms could be impacted (Jensen et Kletzer 2010). The result is that the potential of this transformation is sizeable. However, little has been done on understanding and documenting the enablers and consequences of this journey. If this article does not address what is needed to walk the talk in MC firms, it provides a necessary addition to the field by describing what the length and the end of the journey looks like.

This discussion is articulated around two elements. First, a competitive market where MC firms displace labor is modeled in section 2. This generates the previously described results around the price and profitability dynamics of consulting services. The viability of the model is then illustrated in section 3. The associated proof of concept is built on MC firms dedicated to the support of the German transportation industry and is enabled by public data curated by the OECD. This example, albeit arbitrary at first glance, has been chosen based on two reasons. On one hand, globalization is mainly about displacing work from mature markets to emerging ones. Germany, as the third largest mature market for consulting services in the world (after the US and the UK), thus appears as an appropriate example. On the other hand, management consulting firms are very specialized (Kipping et Engwall 2002). Given that they all aim at providing solutions to their client to improve their performance, they differentiate themselves by tailoring their offerings to industrial sectors. The transportation industry, with its specific challenges, is a representative instance of a typical market for consulting firms and considering its prominence in Germany (as well as other countries), it looks to be a relevant test case for this type of model.

## 2. A model of competition between MC firms in a globalized environment:

### a. Aligning services demand & supply through prices:

For a client firm, contracting a consulting service to improve its performance yields benefits. According to (Armbrüster 2006), those benefits increase with clients' size. For the sake of simplicity, let us assume that this relationship is linear (i.e. consulting benefits are directly proportional (up to factor $v \geq 0$) to client's revenue $r$[2]). A client will only use a service if its benefits ($v.r$) outweigh the associated cost, represented by the service price $p$ (i.e. $v.r \geq p$). Thus, calling $f(t,r)$ the number of firms at time t with revenue $r$, the demand for consulting services $D(t,p)$ can be expressed as:

$$D(t,p) = \int_{r \geq \frac{p}{v}} f(t,r).dr$$

To meet this demand, MC providers do not rely on technology but on workers (Løwendahl 2005). Given that consulting offerings are usually complex, their production for one client requires multiple consultants (say $n$) with different types/level of skills (Maister 2012). A MC practice with a total of e workers can therefore service $\frac{e}{n}$ client firms. Considering this production structure, MC practices have a natural interest in lowering their labor cost to increase their competitiveness. This can be achieved by localizing resources in the right place (e.g. for the same level of skill, a consultant in New York - US will not bear the same cost as a consultant in Johannesburg – South Africa). However, not all firms are equal when it comes to the leveraging geographically disperse resources. It is indeed easier for large firms (especially multi-national ones) to reallocate work across their markets and offices compared to small practices, which are only present in a handful of locations. These differences can be accounted for by assuming that the portion $\phi(e)$ of labor that can be done in a remote location is proportional to the provider size [to a factor $\beta$ (i.e. $\phi(e) = \max(\beta.e, 1)$)].

MC firms only choose to supply services at a price $p$ if their production activities yield a profit. Since local resources come at a unit cost $c$ and displacing a percent of labor off/near-shore yields a savings $\Delta c$, the price of an offering is subject to two conditions. First, services can only be provided if they are profitable when fully produced remotely (i.e. $p + n.\Delta c > n.c$). Second, a MC firm with $e$ workers (a $\phi(e)$ portion of which is remote) can only deliver services at a price $p$ that is above its production costs. This means that only practices of size $e \geq \frac{n.c-p}{n.\beta.\Delta c}$ can deliver a service for a price p. Therefore, calling $g(t,e)$ the number of MC firms with $e$ employees, the supply $S(t,p)$ of service that can be provided at price $p$ is given by:

---

[2] Note that in the economic and management literature, firm size is usually assessed either through its revenue (e.g. (Brush, Bromiley et Hendrickx 2000)) or its number of employees (see (Pagano et Schivardi 2003) for an example).

$$\forall\, p > (c - \Delta c);\quad S(t,p) = \frac{1}{n} \cdot \int_{e \geq \max\left(\frac{n.c-p}{n.\beta.\Delta c};0\right)} e \cdot g(t,e)\, de$$

In this market structure, at any point in time t, the equilibrium price $P(t)$ for MC services therefore adjusts itself so that the available supply of services matches its demand:

$$\forall t,\quad D(t,p) = S(t,p) \leftrightarrow \int_{r \geq \frac{P(t)}{v}} f(t,r)\, dr = \frac{1}{n} \cdot \int_{e \geq \max\left(\frac{n.c-P(t)}{n.\beta.\Delta c};0\right)} e \cdot g(t,e)\, de$$

### b. What are the key dynamics that drive MC prices evolution?

To understand how this equilibrium price $P(t)$ of MC services evolves over time, the dynamics behind the demand and supply in consulting services must be specified. On the demand side, firms' dynamics can be simply represented by Gibrat's law (Sutton 1997) (i.e. clients grow their revenue over time at a constant rate $\psi$ [$i.e.\ dr = \psi.r.dt$]). Considering that a number $h(t)$ of new firms enters the market at time t with a revenue $r_m$ (which is sufficient for them be profitable), demand dynamics, which are exogenous to the consulting industry, are given by:

$$\forall r \geq r_m,\quad \partial_t f + \partial_r(\psi.r.f) = 0 \leftrightarrow f(t,r) = h\left(t - \frac{1}{\psi}.\ln\left(\frac{r}{r_m}\right)\right)$$

The same type of model can also be used on the supply side of the market. As noted in (Audretsch, et al. 2004), consulting firms indeed grow their workforce at a constant pace $\mu$ (i.e. $de = \mu.e.dt$), which represents their ability to source and train young graduates into consultants (Maister 2012). MC firms, which employ at least n workers locally, have therefore their dynamics depicted by:

$$\forall t,\quad \forall e \geq n\ \ \partial_t g + \partial_e(\mu.e.g) = 0 \leftrightarrow g(t,e) = g\left(t - \frac{1}{\mu}.\ln(e), n\right)$$

In this type of set up, two types of markets (and price equilibriums) appear (see figure 1).

**Proposition 1.** A emerging consulting market is a market where the change in consulting demand between two periods of time is higher than the increase in supply due to consulting firm training capabilities (i.e. $\psi.n.\frac{c-\beta.\Delta c}{v}.f\left(t,n.\frac{c-\beta.\Delta c}{v}\right) \geq \mu.D(t,\grave{}n.(c-\beta.\Delta c))$ ). In this type of market, the price of consulting is $P(t) = n.(c - \beta\Delta c)$ and the gap between supply and demand triggers the entry of $g(t,n)$ new firms, such that:

$$g(t,n) = \psi.n.\frac{c-\beta.\Delta c}{v}.f\left(t,\frac{c-\beta.\Delta c}{v}\right) - \mu.D(t,\grave{}n.(c-\beta.\Delta c))$$

**Proposition 2.** A mature consulting market is characterized by an excess of supply over demand (i.e. $\psi.\frac{P(t)}{v}.f\left(t,\frac{P(t)}{v}\right) \geq \mu.D(t,\grave{}P(t))$). This type of market regulates itself through price reductions ($\partial_t P < 0$). Entry is no longer an option and the $k(t) = g\left(t,\frac{n.c-P(t)}{n.\beta.\Delta c}\right).\frac{\partial_t P}{n.\beta.\Delta c}$ least competitive MC practices are exiting the market. Price then adjust to ensure that the increase in the number of profitable providers met the increase in clients:

$$\partial_t P.g\left(t,\frac{n.c-P(t)}{n.\beta.\Delta c}\right).\left(\frac{1}{n.\beta.\Delta c}\right) = f\left(t,\frac{P(t)}{v}\right).\left(\psi.\frac{P(t)}{v}\right) - \mu.S(t,P(t))$$

Note that, at an aggregated level, price reductions come with an increased reliance on labor displacement as leveraging only local workers to deliver service is not a profitable/viable option.

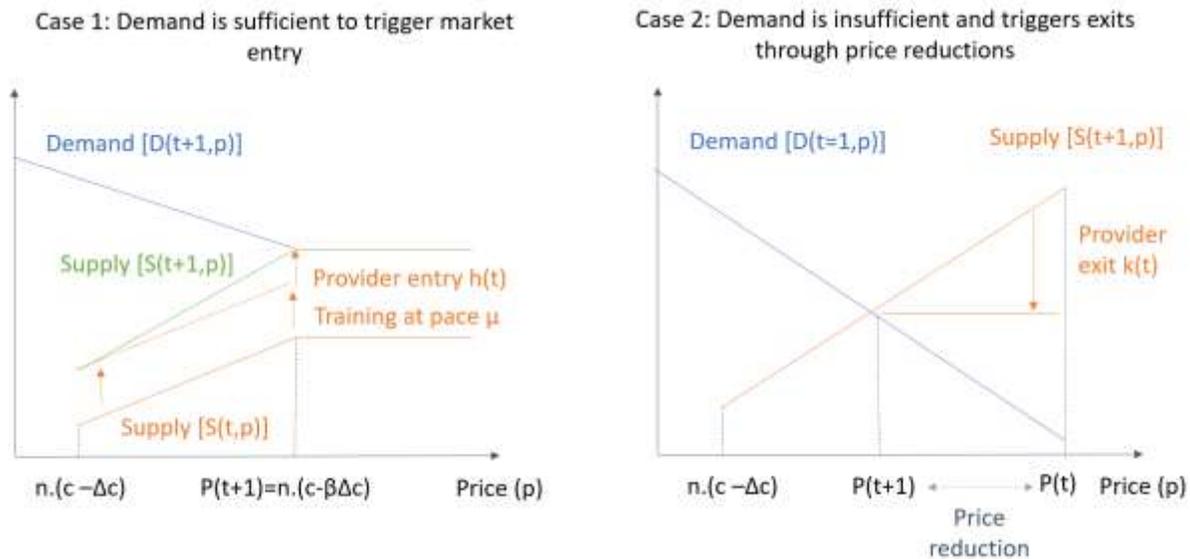

*Figure 1 - Equilibrium between demand and supply in emerging and mature consulting markets.*

Finally, this competitive equilibrium can yield two outcomes on MC firms' profitability. When demand increases enough to trigger entry, the profitability of an individual MC firm of size $e$ increases. Prices and revenues indeed stay flat while the firm keeps transforming its service delivery model (through labor displacement), therefore reducing its production costs. However, when demand dynamics are not sufficient to sustain the entry of new providers, MC firms profitability increases only if they are large enough for their transformation to offset the prices decrease triggered by the market (i.e. $\beta.\mu.e.n.\Delta c \geq \partial_t P(t)$).

## 3. Model application to the German transportation market.

To illustrate its properties, an application of the model to consulting services provided to the German transportation and storage sector is now proposed.

### a. Data:

On the demand side, the calibration of the model can be achieved through the data provided by the OECD[3]. The organization's records show that, over the past 10 years, the transportation and storage sector in Germany (identified through the ISIC4 code 49 to 53) has increased its revenue by about 11.1 ($\pm$ 4.3) $B€$ a year and had an annual birth rate of $\frac{h(t)}{\int f(t,r)dr} \approx 7.3\%$ ($\pm 0.9\%$) . Each new entrant generated each about $r_m \approx 1.3 M€$ ($\pm 0.1$) of yearly revenue. The revenue evolution of the sector displayed on the records therefore translated, as per the proposed model, in an average growth rate for firms is of $\psi \approx 3.6\%$ ($\pm 1.4\%$).

---

[3] https://data.oecd.org/

When it comes to clients' appetite for MC services, one must recall that consulting practices mainly help firms optimize their selling, general and administrative costs, which are worth about 25 to 30% of their revenue (see (Chen, Lu et Sougiannis 2012) or (Anderson, et al. 2007) for benchmarks). So, in the case of a consulting solution that can help firms decrease their general costs by 10%, client generate savings worth around $v \approx 2{,}5\%$ of their revenue.

The supply of those services rely on local German workers, who, according to Glassdoor[4], cost around $c \approx 50k€/year$ and which activities can potentially be displaced in a nearshore hub, say for example in Poland, where salaries are about $\frac{\Delta c}{c} = 50\%$ lower. At peak, the intensity of this work reallocation is worth about 20% (Jensen et Kletzer 2010) and recent empirical studies (see (Metters 2008) for instance) have shown that this peak is only achieved in very large firms (i.e. more than 1000+ employees). This means that for every employee, a MC practice can potentially displace $\beta \approx 0.02\%$ of its total volume of activity.

### b. Results:

With the proposed framework, assuming a constant birth rate $\alpha = \frac{h(t)}{\int f(t,r)dr}$ for potential clients' firms leads to the following demand structure:

$$D(t,p) = F_0 . \alpha . e^{\alpha . t} . \left(\frac{1}{r_m}\right)^{-\frac{\alpha}{\psi}} . \frac{\psi}{\alpha - \psi} . \left(\frac{p}{v}\right)^{\frac{\psi - \alpha}{\psi}} \quad ; \quad f(t,r) = \alpha . F_0 . e^{\alpha . t} . \left(\frac{r}{r_m}\right)^{-\frac{\alpha}{\psi}}$$

For the increase in demand to enable the birth of new consulting practices, the training capabilities of those practices must be such that: $\alpha \geq \mu$. Given the order of magnitudes recovered from the OECD data bank, this would mean that $\mu \leq 3.7\%$. However, employment in the MC field has been growing at a 7%+ rate on average over the past 10 years. It therefore seems unlikely that the increase in demand from the transportation sector would be enough to sustain the formation of new practices. This means that the supply consulting services to the German transportation sector is prone to a competition and a downwards pressure on prices.

To estimate the price pressure in the German market, its supply curve must be built. Since the OECD records associated to German MC firms indicate that they currently follow a pattern similar to (Axtell 2001) of a Zipf distribution in their size ($g(0,e) = \frac{g_0}{e}$), the proposed framework translates into the following supply structure:

$$\forall p \leq P(t), S(t,p) = \frac{e^{\mu t} . g_0}{n . \beta} . (1 + \frac{p - n.c}{n.\Delta c})$$

At that stage, some additional assumptions are needed since the OECD databank is not granular enough to explain how many MC firms service the transportation sector. To deliver the solution, assume, that a total of $n = 1$ worker is required per client. Additionally, assume that only the 7500 largest firms in the transportation sector are currently able to afford the desired service. This translates into $g_0 \approx 16$, an equilibrium price of $P(0) = 37k€$ and requires at least 48% of activities to be off/near shored (see figure 2).

---

[4] https://www.glassdoor.com/index.htm

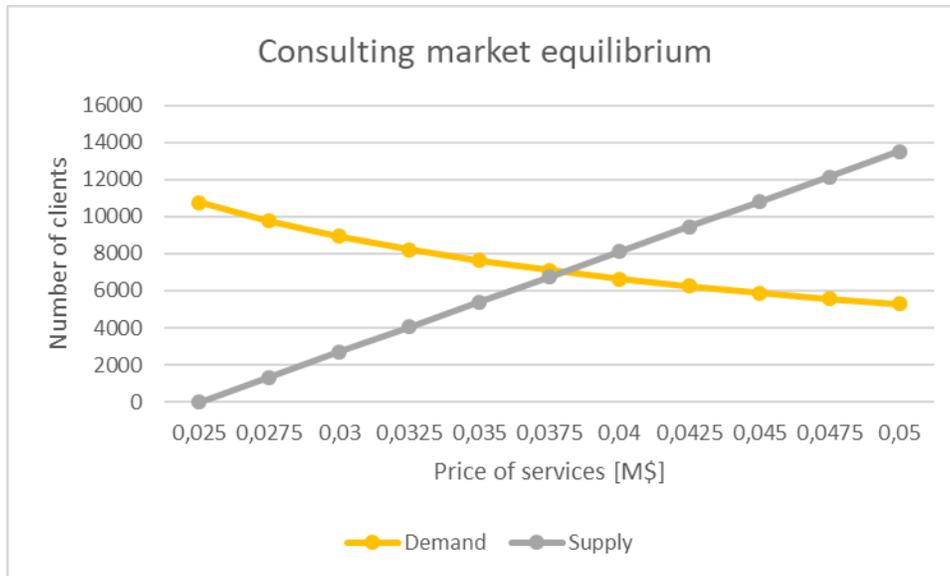

*Figure 2 - MC equilibrium*

Finally running the model leads to the dynamics observed in figure 3. Prices go down by -0.5% a year, while the proportion of off/near-shore work which is required to be remain competitive increase by 0.7% on a year on year basis.

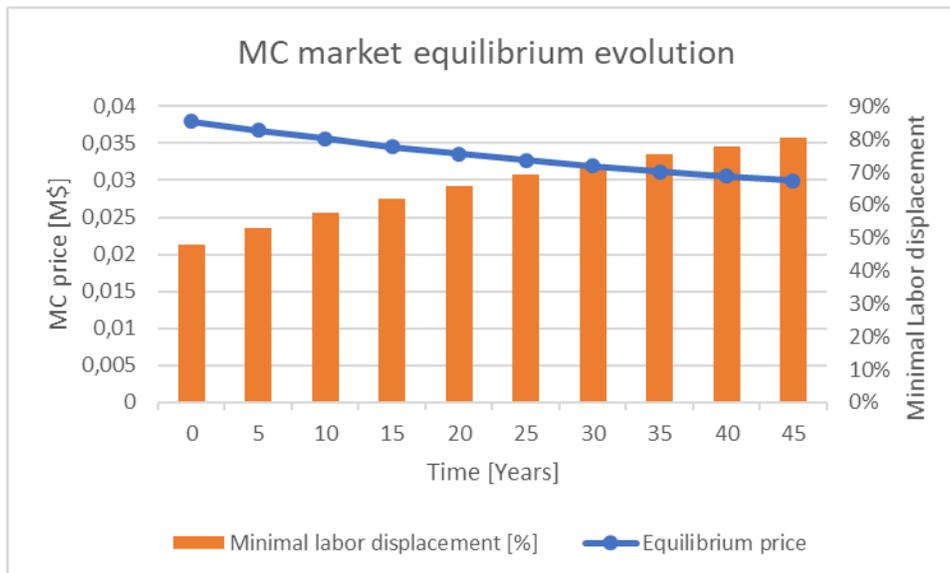

*Figure 3 - MC market evolution*

## 4. Conclusion

This article develops a model showcasing the effect of labor displacement on the management consulting sector. In the case of the German market and the consulting services dedicated to the transportation industry, it shows some consolidation is already ongoing, which results in price reductions of -0.5% a year. In such a mature market, the only firms that can survive over the long run are the ones that improve their competitiveness by pushing at least an extra +0.7% of labor off-near shore every year.